\pdfoutput=1

\documentclass[12pt,a4paper]{article}

\usepackage{ifthen} 
\newboolean{pdflatex}
\setboolean{pdflatex}{true} 

\newboolean{articletitles}
\setboolean{articletitles}{true} 

\newboolean{uprightparticles}
\setboolean{uprightparticles}{false} 

\newboolean{inbibliography}
\setboolean{inbibliography}{false} 

\def\paperauthors{LHCb collaboration} 
\def\paperasciititle{Measurement of the branching fraction and CP asymmetry in B+ -> J/psi rho+ decays} 
\def\papertitle{Measurement of the branching fraction and \CP asymmetry in \decay{\Bp}{\jpsi \rhop} decays} 
\def\paperkeywords{{High Energy Physics}, {LHCb}} 
\def\papercopyright{\the\year\ CERN for the benefit of the LHCb collaboration} 
\def\paperlicence{CC-BY-4.0 licence}
\def\paperlicenceurl{https://creativecommons.org/licenses/by/4.0/}

\usepackage[textwidth=2cm,colorinlistoftodos]{todonotes}

\usepackage[top=1in, bottom=1.25in, left=1in, right=1in]{geometry}

\usepackage{microtype}
\usepackage{lineno}  
\usepackage{xspace} 
\usepackage{caption} 

\usepackage{graphicx}  
\usepackage{color}
\usepackage{colortbl}
\graphicspath{{./figs/}} 

\usepackage{amsmath} 
\usepackage{amssymb}
\usepackage{amsfonts}
\usepackage{upgreek} 

\newcommand*\patchAmsMathEnvironmentForLineno[1]{%
\expandafter\let\csname old#1\expandafter\endcsname\csname #1\endcsname
\expandafter\let\csname oldend#1\expandafter\endcsname\csname
end#1\endcsname
 \renewenvironment{#1}%
   {\linenomath\csname old#1\endcsname}%
   {\csname oldend#1\endcsname\endlinenomath}%
}
\newcommand*\patchBothAmsMathEnvironmentsForLineno[1]{%
  \patchAmsMathEnvironmentForLineno{#1}%
  \patchAmsMathEnvironmentForLineno{#1*}%
}
\AtBeginDocument{%
\patchBothAmsMathEnvironmentsForLineno{equation}%
\patchBothAmsMathEnvironmentsForLineno{align}%
\patchBothAmsMathEnvironmentsForLineno{flalign}%
\patchBothAmsMathEnvironmentsForLineno{alignat}%
\patchBothAmsMathEnvironmentsForLineno{gather}%
\patchBothAmsMathEnvironmentsForLineno{multline}%
\patchBothAmsMathEnvironmentsForLineno{eqnarray}%
}

\usepackage{hyperxmp}

\usepackage[pdftex,
            pdfauthor={\paperauthors},
            pdftitle={\paperasciititle},
            pdfkeywords={\paperkeywords},
            pdfcopyright={Copyright (C) \papercopyright},
            pdflicenseurl={\paperlicenceurl}]{hyperref}

\usepackage[all]{hypcap} 

\usepackage{xspace} 
\usepackage{upgreek}

\def\lhcb {\mbox{LHCb}\xspace}

\def\babar  {\mbox{BaBar}\xspace}

\def\MagUp {\mbox{\em Mag\kern -0.05em Up}\xspace}

\ifthenelse{\boolean{uprightparticles}}%
{
 
 \def\Pgamma      {\ensuremath{\upgamma}\xspace}

 \def\Pmu         {\ensuremath{\upmu}\xspace}

 \def\Ppi         {\ensuremath{\uppi}\xspace}                 
                  
 \def\Prho        {\ensuremath{\uprho}\xspace}

 \def\Ppsi        {\ensuremath{\uppsi}\xspace}

 \def\PDelta      {\ensuremath{\Delta}\xspace}                 
 \def\PXi      {\ensuremath{\Xi}\xspace}                 
 \def\PLambda      {\ensuremath{\Lambda}\xspace}                 
 \def\PSigma      {\ensuremath{\Sigma}\xspace}                 
 \def\POmega      {\ensuremath{\Omega}\xspace}                 
 \def\PUpsilon      {\ensuremath{\Upsilon}\xspace}                 
 
 \def\PB      {\ensuremath{\mathrm{B}}\xspace}                 
                  
 \def\PD      {\ensuremath{\mathrm{D}}\xspace}

 \def\PJ      {\ensuremath{\mathrm{J}}\xspace}                 
 \def\PK      {\ensuremath{\mathrm{K}}\xspace}

 \def\Pb      {\ensuremath{\mathrm{b}}\xspace}                 
 \def\Pc      {\ensuremath{\mathrm{c}}\xspace}                 
 \def\Pd      {\ensuremath{\mathrm{d}}\xspace}

 \def\Pi      {\ensuremath{\mathrm{i}}\xspace}

 \def\Ps      {\ensuremath{\mathrm{s}}\xspace}

}
{
 
 \def\Pgamma      {\ensuremath{\gamma}\xspace}

 \def\Pmu         {\ensuremath{\mu}\xspace}

 \def\Ppi         {\ensuremath{\pi}\xspace}                 
                  
 \def\Prho        {\ensuremath{\rho}\xspace}

 \def\Ppsi        {\ensuremath{\psi}\xspace}                 
                  
 \mathchardef\PDelta="7101
 \mathchardef\PXi="7104
 \mathchardef\PLambda="7103
 \mathchardef\PSigma="7106
 \mathchardef\POmega="710A
 \mathchardef\PUpsilon="7107
                  
 \def\PB      {\ensuremath{B}\xspace}                 
                  
 \def\PD      {\ensuremath{D}\xspace}

 \def\PJ      {\ensuremath{J}\xspace}                 
 \def\PK      {\ensuremath{K}\xspace}

 \def\Pb      {\ensuremath{b}\xspace}                 
 \def\Pc      {\ensuremath{c}\xspace}                 
 \def\Pd      {\ensuremath{d}\xspace}

 \def\Pi      {\ensuremath{i}\xspace}

 \def\Ps      {\ensuremath{s}\xspace}

}

\makeatletter
\ifcase \@ptsize \relax
  \newcommand{\miniscule}{\@setfontsize\miniscule{4}{5}}
\or
  \newcommand{\miniscule}{\@setfontsize\miniscule{5}{6}}
\or
  \newcommand{\miniscule}{\@setfontsize\miniscule{5}{6}}
\fi
\makeatother

\DeclareRobustCommand{\optbar}[1]{\shortstack{{\miniscule (\rule[.5ex]{1.25em}{.18mm})}
  \\ [-.7ex] $#1$}}

\def\mup        {{\ensuremath{\Pmu^+}}\xspace}
\def\mun        {{\ensuremath{\Pmu^-}}\xspace} 
\def\mumu       {{\ensuremath{\Pmu^+\Pmu^-}}\xspace}

\def\g      {{\ensuremath{\Pgamma}}\xspace}

\def\dquark    {{\ensuremath{\Pd}}\xspace}

\def\squark    {{\ensuremath{\Ps}}\xspace}

\def\cquark    {{\ensuremath{\Pc}}\xspace}
\def\cquarkbar {{\ensuremath{\overline \cquark}}\xspace}

\def\bquark    {{\ensuremath{\Pb}}\xspace}

\def\pion   {{\ensuremath{\Ppi}}\xspace}
\def\piz    {{\ensuremath{\pion^0}}\xspace}

\def\pip    {{\ensuremath{\pion^+}}\xspace}
\def\pim    {{\ensuremath{\pion^-}}\xspace}

\def\rhomeson {{\ensuremath{\Prho}}\xspace}
\def\rhoz     {{\ensuremath{\rhomeson^0}}\xspace}
\def\rhop     {{\ensuremath{\rhomeson^+}}\xspace}
\def\rhom     {{\ensuremath{\rhomeson^-}}\xspace}

\def\kaon    {{\ensuremath{\PK}}\xspace}
  \def\Kbar    {{\kern 0.2em\overline{\kern -0.2em \PK}{}}\xspace}

\def\KorKbar    {\kern 0.18em\optbar{\kern -0.18em K}{}\xspace}

\def\Kp      {{\ensuremath{\kaon^+}}\xspace}

\def\KS      {{\ensuremath{\kaon^0_{\mathrm{ \scriptscriptstyle S}}}}\xspace}

\def\Kstarz  {{\ensuremath{\kaon^{*0}}}\xspace}

\def\Kstar   {{\ensuremath{\kaon^*}}\xspace}

\def\Kstarp  {{\ensuremath{\kaon^{*+}}}\xspace}

  \def\Dbar    {{\kern 0.2em\overline{\kern -0.2em \PD}{}}\xspace}

\def\DorDbar    {\kern 0.18em\optbar{\kern -0.18em D}{}\xspace}

\def\B       {{\ensuremath{\PB}}\xspace}
\def\Bbar    {{\ensuremath{\kern 0.18em\overline{\kern -0.18em \PB}{}}}\xspace}

\def\BorBbar    {\kern 0.18em\optbar{\kern -0.18em B}{}\xspace}

\def\Bu      {{\ensuremath{\B^+}}\xspace}
\def\Bub     {{\ensuremath{\B^-}}\xspace}
\def\Bp      {{\ensuremath{\Bu}}\xspace}

\def\Bd      {{\ensuremath{\B^0}}\xspace}
\def\Bs      {{\ensuremath{\B^0_\squark}}\xspace}
\def\Bsb     {{\ensuremath{\Bbar{}^0_\squark}}\xspace}

\def\jpsi     {{\ensuremath{{\PJ\mskip -3mu/\mskip -2mu\Ppsi\mskip 2mu}}}\xspace}

  \def\Y#1S{\ensuremath{\PUpsilon{(#1S)}}\xspace}

\def\Lbar        {{\ensuremath{\kern 0.1em\overline{\kern -0.1em\PLambda}}}\xspace}
\def\LorLbar    {\kern 0.18em\optbar{\kern -0.18em \PLambda}{}\xspace}

\def\BF         {{\ensuremath{\mathcal{B}}}}

\def\BR         {\BF}
\newcommand{\decay}[2]{\ensuremath{#1\!\to #2}\xspace}         

\def\to                 {\ensuremath{\rightarrow}\xspace}

\def\CP                {{\ensuremath{C\!P}}\xspace}

\newcommand{\ACP}{{\ensuremath{{\mathcal{A}}^{\CP}}}\xspace}

\def\BsToJPsiPhi  {\decay{\Bs}{\jpsi\phi}}
\def\BdToJPsiKst  {\decay{\Bd}{\jpsi\Kstarz}}

\def\BuToJPsiKp  {\decay{\Bu}{\jpsi\Kp}}
\def\BuToJPsipip  {\decay{\Bu}{\jpsi\pip}}
\def\BuToJPsiRhop  {\decay{\Bu}{\jpsi\rhop}}
\def\BubToJPsiRhom  {\decay{\Bub}{\jpsi\rhom}}
\def\BuToJPsiKstp  {\decay{\Bu}{\jpsi\Kstarp}}
\def\BdToJPsiRhoz  {\decay{\Bd}{\jpsi\rhoz}}

\def\AT#1     {\ensuremath{A_{\mathrm{T}}^{#1}}\xspace}           

\def\C#1      {\ensuremath{\mathcal{C}_{#1}}\xspace}                       
\def\Cp#1     {\ensuremath{\mathcal{C}_{#1}^{'}}\xspace}                    
\def\Ceff#1   {\ensuremath{\mathcal{C}_{#1}^{\mathrm{(eff)}}}\xspace}        
\def\Cpeff#1  {\ensuremath{\mathcal{C}_{#1}^{'\mathrm{(eff)}}}\xspace}       
\def\Ope#1    {\ensuremath{\mathcal{O}_{#1}}\xspace}                       
\def\Opep#1   {\ensuremath{\mathcal{O}_{#1}^{'}}\xspace}                    

\newcommand{\tev}{\ifthenelse{\boolean{inbibliography}}{\ensuremath{~T\kern -0.05em eV}}{\ensuremath{\mathrm{\,Te\kern -0.1em V}}}\xspace}
\newcommand{\gev}{\ensuremath{\mathrm{\,Ge\kern -0.1em V}}\xspace}
\newcommand{\mev}{\ensuremath{\mathrm{\,Me\kern -0.1em V}}\xspace}
\newcommand{\kev}{\ensuremath{\mathrm{\,ke\kern -0.1em V}}\xspace}
\newcommand{\ev}{\ensuremath{\mathrm{\,e\kern -0.1em V}}\xspace}
\newcommand{\gevc}{\ensuremath{{\mathrm{\,Ge\kern -0.1em V\!/}c}}\xspace}
\newcommand{\mevc}{\ensuremath{{\mathrm{\,Me\kern -0.1em V\!/}c}}\xspace}
\newcommand{\gevcc}{\ensuremath{{\mathrm{\,Ge\kern -0.1em V\!/}c^2}}\xspace}
\newcommand{\gevgevcccc}{\ensuremath{{\mathrm{\,Ge\kern -0.1em V^2\!/}c^4}}\xspace}
\newcommand{\mevcc}{\ensuremath{{\mathrm{\,Me\kern -0.1em V\!/}c^2}}\xspace}

\def\mum  {\ensuremath{{\,\upmu\mathrm{m}}}\xspace}

\def\invfb   {\ensuremath{\mbox{\,fb}^{-1}}\xspace}

\newcommand{\chisq}{\ensuremath{\chi^2}\xspace}

\newcommand{\chisqip}{\ensuremath{\chi^2_{\text{IP}}}\xspace}

\def\gsim{{~\raise.15em\hbox{$>$}\kern-.85em
          \lower.35em\hbox{$\sim$}~}\xspace}
\def\lsim{{~\raise.15em\hbox{$<$}\kern-.85em
          \lower.35em\hbox{$\sim$}~}\xspace}

\def\ptot       {\mbox{$p$}\xspace}
\def\pt         {\mbox{$p_{\mathrm{ T}}$}\xspace}

\def\evtgen     {\mbox{\textsc{EvtGen}}\xspace}

\def\geant      {\mbox{\textsc{Geant4}}\xspace}

\def\photos     {\mbox{\textsc{Photos}}\xspace}

\def\pythia     {\mbox{\textsc{Pythia}}\xspace}

\def\tell1  {TELL1\xspace}
\def\ukl1   {UKL1\xspace}

\usepackage{cite} 
\usepackage{mciteplus}

\usepackage{longtable} 

\begin{document}

\renewcommand{\thefootnote}{\fnsymbol{footnote}}
\setcounter{footnote}{1}


\begin{titlepage}
\pagenumbering{roman}

\vspace*{-1.5cm}
\centerline{\large EUROPEAN ORGANIZATION FOR NUCLEAR RESEARCH (CERN)}
\vspace*{1.5cm}
\noindent
\begin{tabular*}{\linewidth}{lc@{\extracolsep{\fill}}r@{\extracolsep{0pt}}}
\ifthenelse{\boolean{pdflatex}}
{\vspace*{-1.5cm}\mbox{\!\!\!\includegraphics[width=.14\textwidth]{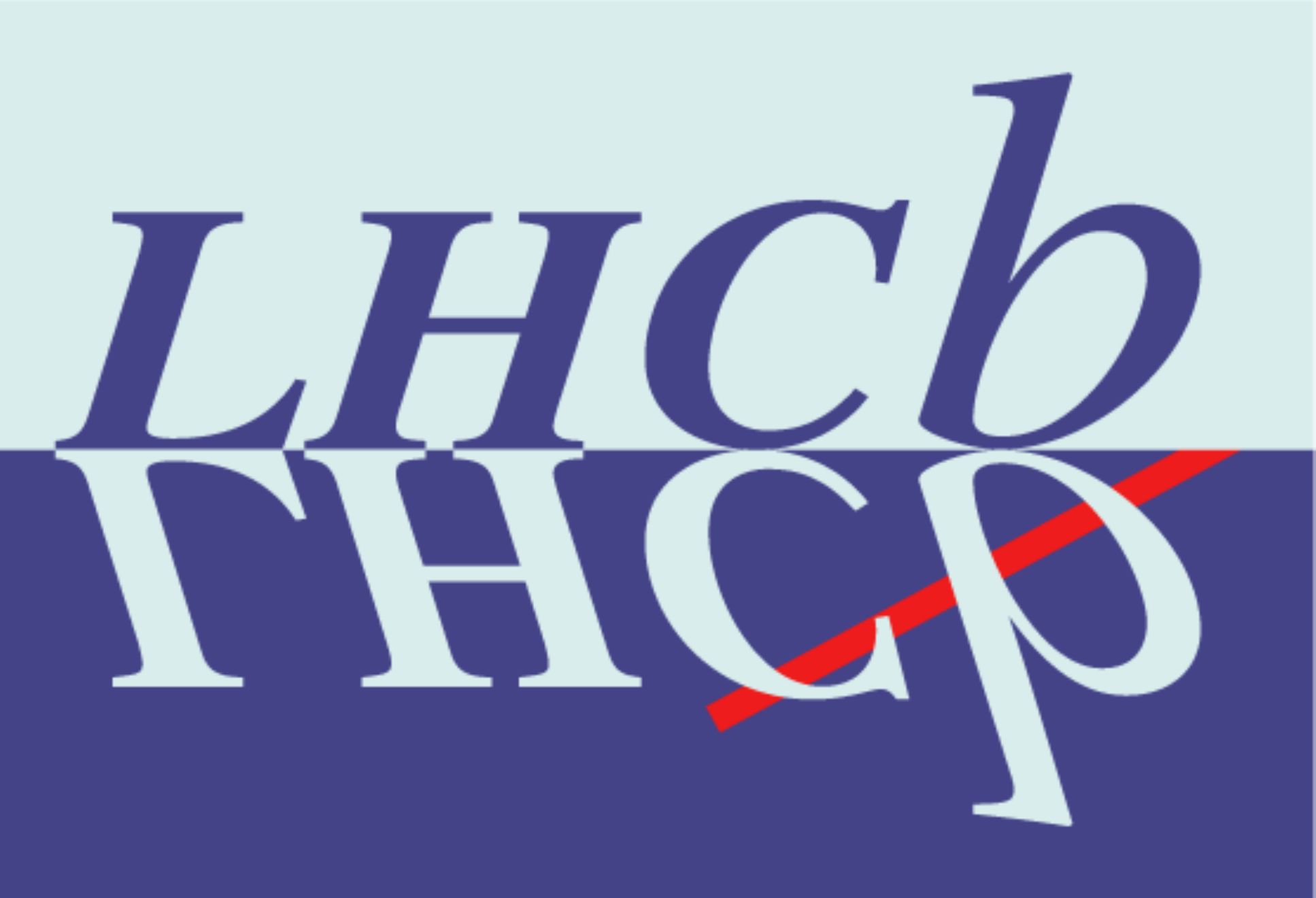}} & &}%
{\vspace*{-1.2cm}\mbox{\!\!\!\includegraphics[width=.12\textwidth]{lhcb-logo.eps}} & &}%
\\
 & & CERN-EP-2018-298 \\  

 & & LHCb-PAPER-2018-036 \\  
 & & June 27, 2019 \\ 
 & & \\
\end{tabular*}

\vspace*{1.0cm}

{\normalfont\bfseries\boldmath\huge
\begin{center}
  \papertitle 
\end{center}
}

\vspace*{1.0cm}

\begin{center}
\paperauthors\footnote{Authors are listed at the end of this paper.}
\end{center}

\vspace{\fill}

\begin{abstract}
  \noindent
 The branching fraction and direct \CP asymmetry of the decay \decay{\Bp}{\jpsi\rhop} are measured using proton-proton collision data collected with the LHCb detector at centre-of-mass energies of 7 and 8 TeV, corresponding to a total integrated luminosity of 3\invfb. 
The following results are obtained:
\begin{align}
\mathcal{B}(\BuToJPsiRhop) &= (3.81 ^{+0.25}_{-0.24} \pm 0.35) \times 10^{-5},  \nonumber \\
\mathcal{A}^{\CP} (\BuToJPsiRhop) &= -0.045^{+0.056}_{-0.057} \pm 0.008, \nonumber 
\end{align}
where the first uncertainties are statistical and the second systematic. Both measurements are the most precise to date.

\end{abstract}

\vspace*{1.0cm}

\begin{center}
  Published in Eur. Phys. J. C\textbf{79} (2019) 537
\end{center}

\vspace{\fill}

{\footnotesize 
\centerline{\copyright~\papercopyright. \href{\paperlicenceurl}{\paperlicence}.}}
\vspace*{2mm}

\end{titlepage}


\newpage
\setcounter{page}{2}
\mbox{~}
%

\cleardoublepage


\renewcommand{\thefootnote}{\arabic{footnote}}
\setcounter{footnote}{0}



\pagestyle{plain} 
\setcounter{page}{1}
\pagenumbering{arabic}


%


\section{Introduction}
\label{sec:Introduction}

\begin{figure}[hb]
  \begin{center}
    \includegraphics[width=0.8\linewidth]{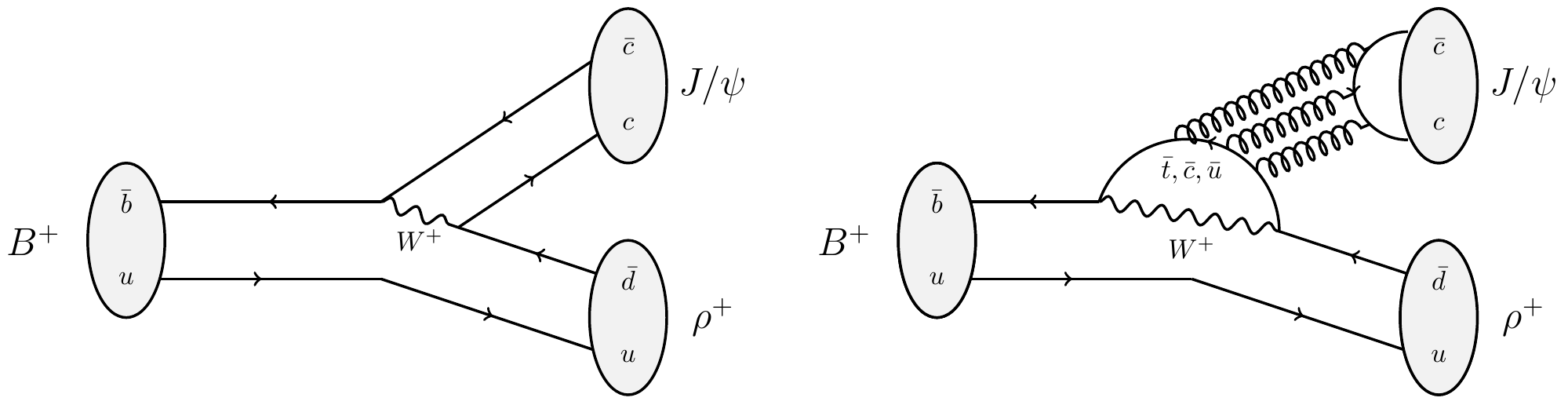}
    \vspace*{-0.5cm}
  \end{center}
  \caption{Leading-order Feynman diagrams for the (left) tree and (right) penguin amplitudes contributing to the decay \BuToJPsiRhop.}
  \label{fig:feynman}
\end{figure}

 In the Standard Model of particle physics, the decay \BuToJPsiRhop proceeds predominantly via a \bquark \to ~\cquark\cquarkbar\dquark transition involving tree and penguin amplitudes,\footnote{Charge conjugation is implied throughout this paper, unless otherwise stated.} as shown in Fig.~\ref{fig:feynman}. Interference between these two amplitudes can lead to direct \CP violation that is measured through an asymmetry defined as 
\begin{equation}
\ACP \equiv \frac{\BR(\BubToJPsiRhom) - \BR(\BuToJPsiRhop)}{\BR(\BubToJPsiRhom) + \BR(\BuToJPsiRhop)}.
\end{equation}
No precise prediction for \ACP exists, though it is expected to have an absolute value ${\lesssim 0.35}$~\cite{PhysRevLett.115.061802} assuming isospin symmetry between the \BdToJPsiRhoz and the \BuToJPsiRhop decays.
Measurements of \ACP provide an estimate of the imaginary part of the penguin-to-tree amplitude ratio for the \bquark \to ~\cquark\cquarkbar\dquark transition. Similarly to the \BdToJPsiRhoz decay \cite{LHCb-PAPER-2014-058}, the \CP asymmetry is expected to be enhanced in this decay compared to the decay \BsToJPsiPhi~\cite{PhysRevD.79.014005,PhysRevD.60.073008}. Therefore its value can be used to place constraints on penguin effects in measurements of the \CP-violating phase $\phi_s$ from the decay \BsToJPsiPhi, assuming approximate SU(3) flavour symmetry and neglecting exchange and annihilation diagrams. 
The branching fraction and the value of \ACP for \decay{\Bp}{\jpsi\rhop} decays
were measured previously by the \babar collaboration to be $(5.0\pm0.7\pm0.3) \times 10^{-5}$ and $-0.11\pm0.12\pm0.08$, respectively~\cite{PhysRevD.76.031101}.

In this paper, the branching fraction and the direct \CP asymmetry of the decay \decay{\Bp}{\jpsi\rhop} are measured using proton-proton $(pp)$ collision data collected with the LHCb detector at centre-of-mass energies of 7 TeV (in 2011) and 8 TeV (in 2012), corresponding to a total integrated luminosity of 3\invfb. The \BuToJPsiRhop decay is analysed using the \decay{\jpsi}{\mumu}, \decay{\rhop}{\pip\piz} and \decay{\piz}{\g\g} decays. Its branching fraction is measured relative to that of the abundant decay \BuToJPsiKp, which has the same number of charged final-state particles and contains a \jpsi meson as the decay of interest.

\section{Detector and simulation}
\label{sec:Detector}
The \lhcb detector~\cite{Alves:2008zz,LHCb-DP-2014-002} is a single-arm forward
spectrometer covering the \mbox{pseudorapidity} range $2<\eta <5$,
designed for the study of particles containing \bquark or \cquark
quarks. The detector includes a high-precision tracking system
consisting of a silicon-strip vertex detector surrounding the $pp$
interaction region~\cite{LHCb-DP-2014-001}, a large-area silicon-strip detector located
upstream of a dipole magnet with a bending power of about
$4{\mathrm{\,Tm}}$, and three stations of silicon-strip detectors and straw
drift tubes~\cite{LHCb-DP-2013-003} placed downstream of the magnet.
The tracking system provides a measurement of the momentum, \ptot, of charged particles with
a relative uncertainty that varies from 0.5\% at low momentum to 1.0\% at 200\gevc.
The minimum distance of a track to a primary vertex (PV), the impact parameter (IP), is measured with a resolution of $(15+29/\pt)\mum$,
where \pt is the component of the momentum transverse to the beam, in\,\gevc.
Different types of charged hadrons are distinguished using information
from two ring-imaging Cherenkov (RICH) detectors~\cite{LHCb-DP-2012-003}. 
Photons, electrons and hadrons are identified by a calorimeter system consisting of scintillating-pad and preshower detectors, an electromagnetic
and a hadronic calorimeter. Muons are identified by a
system composed of alternating layers of iron and multiwire
proportional chambers~\cite{LHCb-DP-2012-002}.

The magnetic field deflects oppositely charged particles in opposite
directions which can lead to charge-dependent acceptance effects. 
The configuration with the magnetic field pointing upwards (downwards) bends positively (negatively) charged particles in the horizontal plane towards the centre of the LHC ring.
Periodically reversing the magnetic field polarity throughout the data taking sufficiently cancels these acceptance effects for the precision of this measurement. Furthermore, the possible difference in material interactions between positively and negatively charged pions is negligible for this analysis~\cite{LHCb-PAPER-2012-009}.

The online event selection is performed by a trigger~\cite{LHCb-DP-2012-004}, which consists of a hardware stage, based on information from the calorimeter and muon systems, followed by a software stage, which applies a full event
reconstruction.
For this analysis, the hardware trigger requires at least one muon with a transverse momentum larger than 1.5\gevc in 2011 and 1.8\gevc in 2012. 
The first stage of the software trigger requires either a muon candidate with a momentum larger than 8\gevc and a transverse momentum larger than 1\gevc, or two muons that form a good quality vertex with an invariant mass greater than 2.7\gevcc.
In the second stage of the software trigger, two particles, identified as being muons, must form a good-quality vertex with an invariant mass compatible with the known \jpsi mass\cite{PDG2018}.

Simulated events are used to study the kinematical properties of the \BuToJPsiRhop and \BuToJPsiKp decays, to study the background contamination and to evaluate the selection efficiencies. In the simulation, $pp$ collisions are generated using
\pythia~\cite{Sjostrand:2006za,*Sjostrand:2007gs} 
 with a specific \lhcb
configuration~\cite{LHCb-PROC-2010-056}.  Decays of hadronic particles
are described by \evtgen~\cite{Lange:2001uf}, in which final-state
radiation is generated using \photos~\cite{Golonka:2005pn}. The
interaction of the generated particles with the detector, and its response,
are implemented using the \geant
toolkit~\cite{Allison:2006ve, *Agostinelli:2002hh} as described in
Ref.~\cite{LHCb-PROC-2011-006}.

\section{Selection}
\label{sec:selection}

The same selection criteria are placed on the \jpsi candidates for both decays, \BuToJPsiRhop and \BuToJPsiKp.  Each \jpsi candidate is composed of two tracks compatible with being muons that form a good quality common vertex significantly  displaced from any reconstructed PV in the event. The invariant mass of this two-track combination must be within $\pm$100\mevcc of the known \jpsi mass\cite{PDG2018}. 

For the \BuToJPsiRhop decay, each \rhop candidate is formed from a charged and a neutral pion. The charged pion is required to have a \chisqip value that is inconsistent with originating from any PV in the event,
where \chisqip is defined as the difference between the vertex-fit \chisq of a given PV reconstructed with and without the particle under consideration. 
The charged pion also needs to have a momentum larger than 3\gevc and a transverse momentum larger than 250\mevc. 
To reject background from \BuToJPsiKstp decays\footnote{Throughout this publication, \Kstar refers to the $K^{*}(892)$ meson.}, with \decay{\Kstarp}{\Kp\piz}, where a kaon is misidentified as a pion, a stringent criterion is placed on the pion-identification quality, which is mainly derived using information from the two RICH detectors. The neutral pion is reconstructed from two well-separated clusters in the electromagnetic calorimeter. It is required to have a \pt larger than 800\mevc. The \rhop candidate must have a transverse momentum larger than 800\mevc and an invariant mass $m_{\pip\g\g}$ between 400\mevcc and 1100\mevcc. 

For the \BuToJPsiKp decay, the \Kp candidate is selected among particles that fulfil the same requirements applied to the charged pion in the \BuToJPsiRhop decay, have a transverse momentum larger than 800\mevc and are identified as a kaons.

The \Bu candidate is formed from the \jpsi candidate and the \rhop or \Kp candidate by requiring that the three reconstructed tracks form a good-quality vertex with significant displacement from the PV. In events with multiple PVs, that with which the \Bu candidate has the smallest \chisqip is chosen. In addition, the \Bu candidate is constrained to originate from the PV using a kinematic fit~\cite{Hulsbergen:2005pu}. In the \BuToJPsiRhop channel, the dimuon invariant mass is constrained to the known \jpsi mass, and when considering the invariant mass of the \Bu candidate, the \piz candidate is additionally constrained to its known mass. This \chisqip value is required to be small enough to be consistent with the \Bu meson having originated from the PV. 

Decays of $b$ hadrons with an additional charged particle are rejected by ensuring that the quality of the \Bu decay vertex significantly degrades when the closest additional track is included in the vertex fit.
Finally, the mass of the \Bu candidate, $m_{\jpsi\pip\piz}$ ($m_{\mup\mun\Kp})$, is required to be between 5100\mevcc and 5700\mevcc (5180\mevcc and 5400\mevcc). 
For the \BuToJPsiKp channel, this selection results in a good signal purity, and no further selection criteria are needed. 

Two vetoes are applied for the \BuToJPsiRhop channel to reject \BuToJPsiKp and \BuToJPsipip decays combined with a random \piz: the mass of the combination of the \jpsi and the charged pion candidate, evaluated under both the pion and kaon mass hypotheses, must be outside a 50\mevcc window around the known \Bu mass.

In addition to the preselection discussed above, a multivariate selection is performed on the \BuToJPsiRhop channel using an artificial neural network from the TMVA package\cite{Hocker:2007ht,TMVA4}, mainly to reduce combinatorial background. The 12 input variables are as follows: the flight distance and direction angle of the \Bu candidate, defined as the angle between its momentum and the vector connecting its primary and decay vertices; the transverse momentum of the \Bu and \rhop candidates; the maximum transverse momentum of the two muons, the \chisqip of the charged pion; the minimum \chisqip of the two muons; the quality of the \Bu candidate vertex; the change in the vertex quality when adding the closest track that is not part of the signal candidate; the quality of a kinematic fit of the full decay chain; the quality of the \piz identification; and the \pt asymmetry in a cone around the flight direction of the \Bu meson, defined as $(\sum_{i} \pt_{i} - \sum_{j} \pt_{j})/(\sum_{i} \pt_{i} + \sum_{j} \pt_{j})$, where $i$ runs over all final state tracks of \BuToJPsiRhop and $j$ over all other tracks in a cone around the \Bu meson flight direction. 
The classifier is trained using background events from both a lower (4800--5000\mevcc) and upper (5700--6000\mevcc) sideband of the \Bu candidate mass and  simulated signal decays, where the \pt distribution of the \Bu meson and the number of tracks per event are weighted to match the corresponding distributions in a \BdToJPsiKst data sample, where \decay{\Kstarz}{\Kp\pim}. The \BdToJPsiKst candidates are obtained using the same preselection on the charged particles as for the \BuToJPsiRhop decay, except that a looser particle-identification criterion is used for the pion. Moderate criteria are placed on the kaon $p$, \pt  and particle identification to obtain a good signal purity. For quantities depending on the kinematics of the \piz meson, the channel \BuToJPsiKstp, with \decay{\Kstarp}{\Kp\piz}, is used to compare the distributions of the input variables between simulation and data. For all variables good agreement is found.

The neural network selection criterion is optimized by maximizing the figure of merit $N_\text{sig}/\sqrt{N_\text{sig}+N_\text{bg}}$, where $N_\text{sig}$ is the expected number of signal decays and $N_\text{bg}$ is the estimated background yield, both between 5200\mevcc and 5450\mevcc. The value of $N_\text{sig}$ is calculated using the ratio of the previously measured \BuToJPsiRhop and \BuToJPsiKp branching fractions\cite{PDG2018}, the observed number of \BuToJPsiKp decays, and the efficiencies of the \BuToJPsiRhop and \BuToJPsiKp channels. The value of $N_\text{bg}$ is obtained by extrapolating the shape of the background into the signal region. 
The optimized cut rejects 99.4\% of background events in both, the 2011 and 2012 data samples, while retaining 49\% (45\%) of signal events in the 2011 (2012) data samples.

After applying the full selection for \BuToJPsiRhop decays, about 9\% of events have more than one candidate. Most of these events contain a genuine \BuToJPsiRhop decay, along with a candidate comprised of the charged particles from the signal decay combined with a prompt \piz meson. 
The latter constitutes a peaking background that is difficult to model, and therefore, events with multiple candidates are removed from the analysis. The efficiency of this rejection is evaluated on simulated samples. The sample of \BuToJPsiKp decays contains only 0.2\% events with multiple candidates and no rejection is required.

\section{Efficiencies}
\label{sec:efficiencies}

The branching ratio of the decay \BuToJPsiRhop is calculated using

\begin{equation}
\label{eq:master}
\mathcal{B}(\BuToJPsiRhop) = \mathcal{B}(\BuToJPsiKp) \times \frac{N_{\BuToJPsiRhop}}{N_{\BuToJPsiKp}} \times \frac{\varepsilon_{\BuToJPsiKp}}{\varepsilon_{\BuToJPsiRhop}} \times \frac{1}{\mathcal{B}(\decay{\piz}{\gamma\gamma})},
\end{equation}
with $N_{\BuToJPsiRhop}$ ($N_{\BuToJPsiKp}$) the number of measured \BuToJPsiRhop (\BuToJPsiKp) decays and $\varepsilon_{\BuToJPsiRhop}$ ($\varepsilon_{\BuToJPsiKp}$) the efficiency for the  \BuToJPsiRhop (\BuToJPsiKp) channel. The efficiencies $\varepsilon_{\BuToJPsiKp}$ and $\varepsilon_{\BuToJPsiRhop}$ are composed of the geometrical acceptance, trigger, reconstruction, particle identification and selection efficiencies. In addition, there is an efficiency due to the removal of multiple candidates in the \BuToJPsiRhop sample. 
The efficiency for the decay products of a \BuToJPsiRhop or \BuToJPsiKp decay to be within the acceptance of the \lhcb detector is taken from simulation. The efficiency to trigger on one or both muons from the \jpsi decay is also taken from simulation. When forming the ratio of branching fractions between the \BuToJPsiRhop and \BuToJPsiKp decays, the ratio of trigger efficiencies is close to unity. 

The charged-particle reconstruction efficiency is taken from simulation with kinematics-dependent correction factors applied that are determined using a tag-and-probe technique applied on a detached \decay{\jpsi}{\mumu} data sample~\cite{LHCb-DP-2013-002}. The \piz reconstruction efficiency is also obtained from simulation, with a \pt-dependent correction factor obtained from the deviation between the known and observed ratio of branching fractions of \BuToJPsiKp and \BuToJPsiKstp decays, with \decay{\Kstarp}{\Kp\piz} followed by \decay{\piz}{\gamma\gamma}\cite{LHCb-PAPER-2012-022}. Data samples of \BuToJPsiKp and \BuToJPsiKstp decays, collected in 2011 and 2012 by the \lhcb experiment, are fitted to obtain the number of observed decays. After correcting for the charged-particle selection efficiencies, the double ratio between the ratio of observed decays and the ratio of known branching fractions is the efficiency to reconstruct a \piz meson. This result is compared to the \piz efficiency obtained from simulated samples to determine the correction factors in intervals of the \pt of the \piz meson. Using this procedure, the uncertainty on the branching fraction of the \BuToJPsiKp decay cancels in the measurement of the \BuToJPsiRhop branching fraction, and only the uncertainty on the branching fraction of the \BuToJPsiKstp decay contributes.

The charged-particle identification efficiency is evaluated using a tag-and-probe technique on dedicated calibration data samples with clean signatures \cite{LHCb-PUB-2016-021}. Given the similar kinematics for the muons from \BuToJPsiRhop and \BuToJPsiKp decays, the muon identification efficiency fully cancels when forming the ratio of branching fractions. The charged-pion identification efficiency is 72\% for the 2011 data and 74\% for the 2012 data. The efficiencies for identifying positively and negatively charged pions are compatible within their statistical uncertainties. 
The efficiency for identifying the kaon is above 95\% for both data taking periods. 

The remaining offline selection efficiencies are taken from simulation, where all kinematic distributions for the signal decay are found to be compatible with the corresponding distributions observed in data for \BdToJPsiKst and \BuToJPsiKstp decays.

\section{Invariant mass fits}
\label{sec:fits}

\begin{figure}[t]
  \begin{center}
    \includegraphics[width=0.49\linewidth]{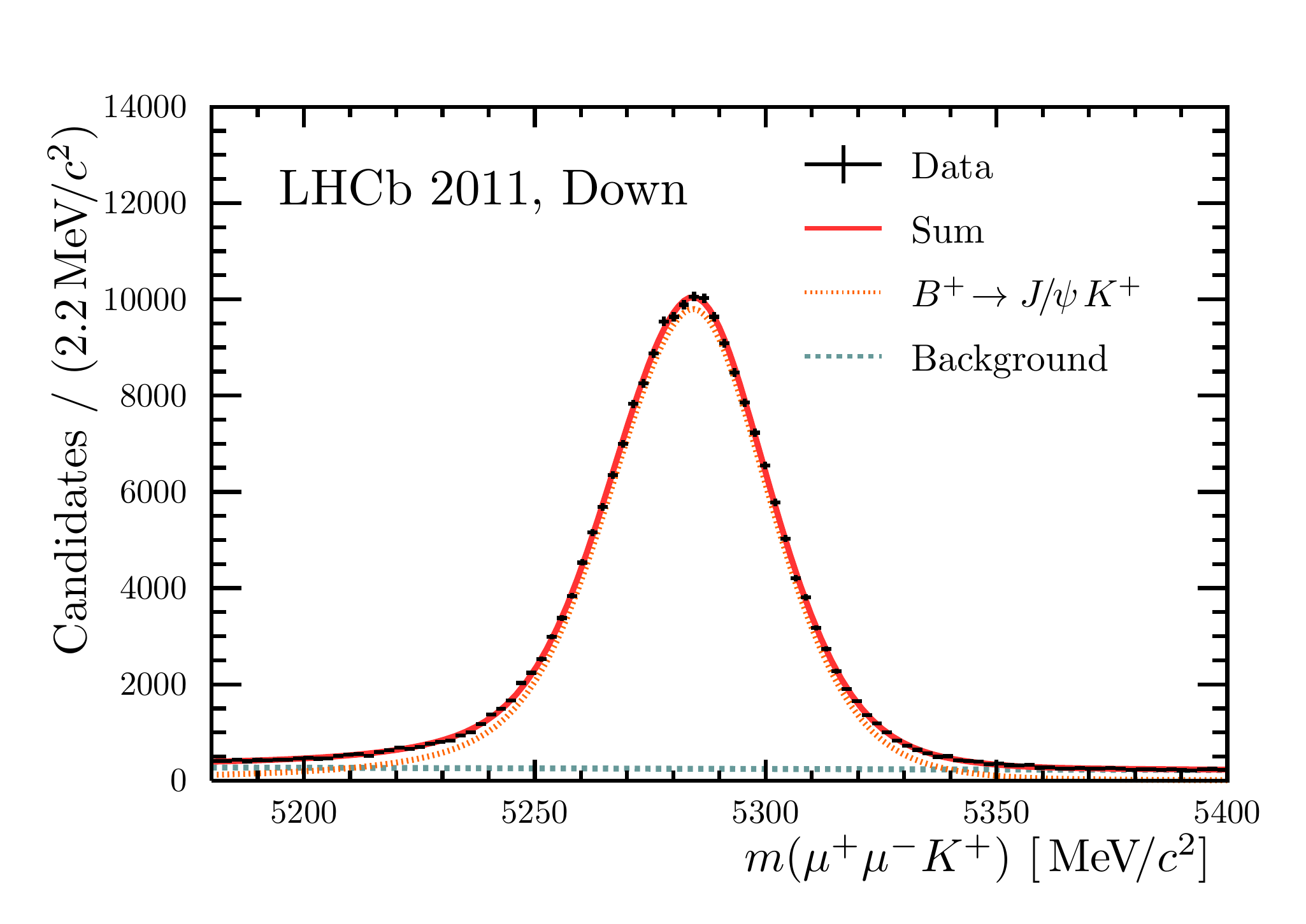}
    \includegraphics[width=0.49\linewidth]{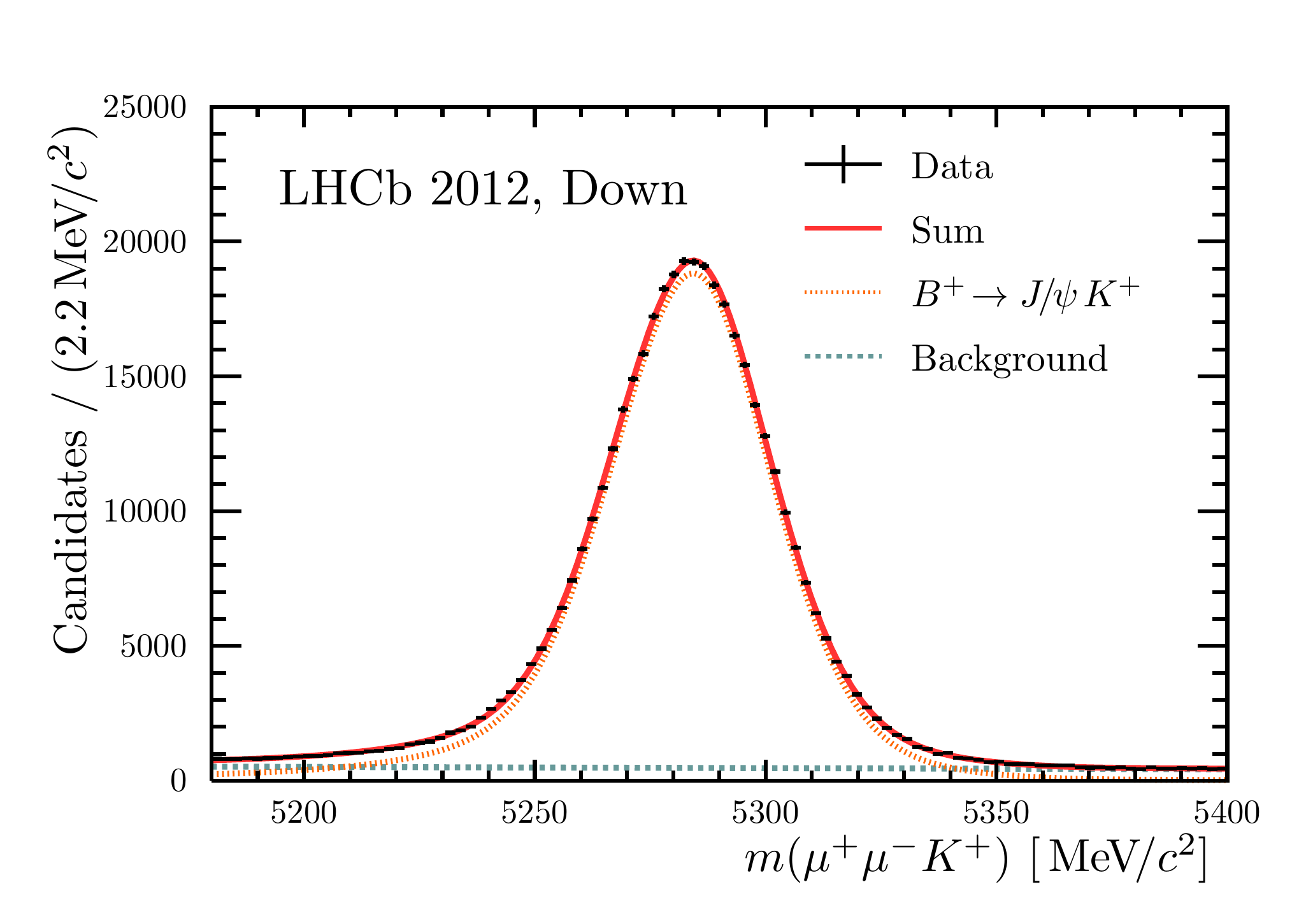}
    \vspace*{-0.5cm}
  \end{center}
  \caption{ The \mup\mun\Kp invariant mass distributions for the (left) 2011 and (right) 2012 data sets with down magnet polarity.}
  \label{fig:JpsiKFit}
\end{figure}

The yield of \BuToJPsiKp decays is determined using an extended unbinned maximum-likelihood fit to the $m_{\mup\mun\Kp}$ distribution.  The signal shape is described using the sum of two Crystal Ball functions~\cite{Skwarnicki:1986xj} and a Gaussian function, where all three functions share the peak position value.  The tail parameters of the Crystal Ball functions are fixed to the values obtained from simulation. An exponential function is used to model the combinatorial background. The fit yields a total of 362\,739 $\pm$ 992 \BuToJPsiKp decays in 2011 and 816\,197 $\pm$ 1545 in 2012. The fit is performed separately for the two magnet polarities; Figure~\ref{fig:JpsiKFit} shows the fit to the data taken with down polarity.

The yield of \BuToJPsiRhop decays is determined using a simultaneous two-dimensional extended unbinned maximum-likelihood fit to the $m_{\jpsi\pip\piz}$ and $m_{\pip\g\g}$ distributions in the 2011 and 2012 data sets. 
The signal shape in $m_{\jpsi\pip\piz}$ is modelled with the sum of two Crystal Ball functions with a shared peak position value. The values of the tail parameters are taken from simulation. The signal shape in $m_{\pip\g\g}$ is modelled with a relativistic Breit--Wigner function

\begin{equation}\mathcal{P}_\text{\rhop}(m_{\pip\g\g}) = \frac{m_{\pip\g\g}\Gamma(m_{\pip\g\g}) P_{\jpsi}^{2L_{\mathrm{eff}}+1} }{(m_{\rhop}^2 - m_{\pip\g\g}^2)^2 + m_{\rhop}^2 \Gamma(m_{\pip\g\g})^2},\end{equation}
with 
\begin{equation}
\Gamma(m_{\pip\g\g}) = \Gamma_0 \left(\frac{q}{q_0}\right)^3 \left(\frac{m_\rhop}{m_{\pip\g\g}}\right) \left(\frac{1+R^2q_0^2}{1+R^2 q^2}\right). 
\end{equation}
Here, $P_{\jpsi}$ is the momentum of the \jpsi meson in the \Bp rest frame, $q(m_{\pip\g\g})$ is the pion momentum in the dipion rest frame, $L_{\mathrm{eff}}$ is the relative angular momentum of the \rhop meson with respect to the \jpsi, $q_{0} = q(m_{\rhop})$, $\Gamma_{0}$ is the nominal width, and $R$ is a barrier factor radius. 
The tail parameters of the Crystal Ball function, along with $m_{\rhop}$ and $\Gamma_{0}$ of the Breit--Wigner, are fixed to values obtained from a fit to simulated \BuToJPsiRhop decays, generated using the same functional form with input values $m_{\rhop} = 768.5\mevcc$ and $\Gamma_{0}$ = 151\mevcc.
This strategy therefore folds in the effect of the $m_{\pip\g\g}$ resolution into the Breit--Wigner model, which results in a value of $\Gamma_{0} = 175\mevcc$.
The relative angular momentum $L_{\text{eff}}$ is fixed to 0.

Non-resonant \jpsi\pip\piz decays are described with the same shape as the signal in $m_{\jpsi\pip\piz}$, while in $m_{\pip\g\g}$ a three-body phase-space distribution multiplied by $P_{\jpsi}^{2}$ is used, which is motivated by angular momentum conservation. Given the slowly varying shape in $m_{\pip\g\g}$, no description of the detector resolution is needed. 
The combinatorial background is described by an exponential function in $m_{\jpsi\pip\piz}$ and a first-order polynomial in $m_{\pip\g\g}$.

Two partially reconstructed backgrounds are included in the \BuToJPsiRhop fit. The first is the decay \BuToJPsiKstp, with \decay{\Kstarp}{\KS\pip} and \decay{\KS}{\piz\piz}, where one \piz is not reconstructed. The second is the decay \BsToJPsiPhi, with \decay{\phi}{\pip\pim\piz}, where one charged pion is not reconstructed. As the shapes of both contributions exhibit significant correlations between $m_{\jpsi\pip\piz}$ and $m_{\pip\g\g}$, they are described by two-dimensional kernel density estimators \cite{Cranmer:2000du} with adaptive kernels determined using simulation.

The final component is the background from \BuToJPsiKstp decays, with \decay{\Kstarp}{\Kp\piz}, where the kaon is misidentified as a pion. Despite the stringent particle-identification criterion applied, a small amount of these decays is present in the final sample. The product of two one-dimensional kernel density estimators is used to describe the shape in the two invariant mass distributions. This background yield is fixed relative to that of the \BuToJPsiKstp decay, with \decay{\Kstarp}{\KS\pip}, using the known ratio of the branching fractions\cite{PDG2018}.

\begin{figure}[tb]
  \begin{center}
    \includegraphics[width=0.49\linewidth]{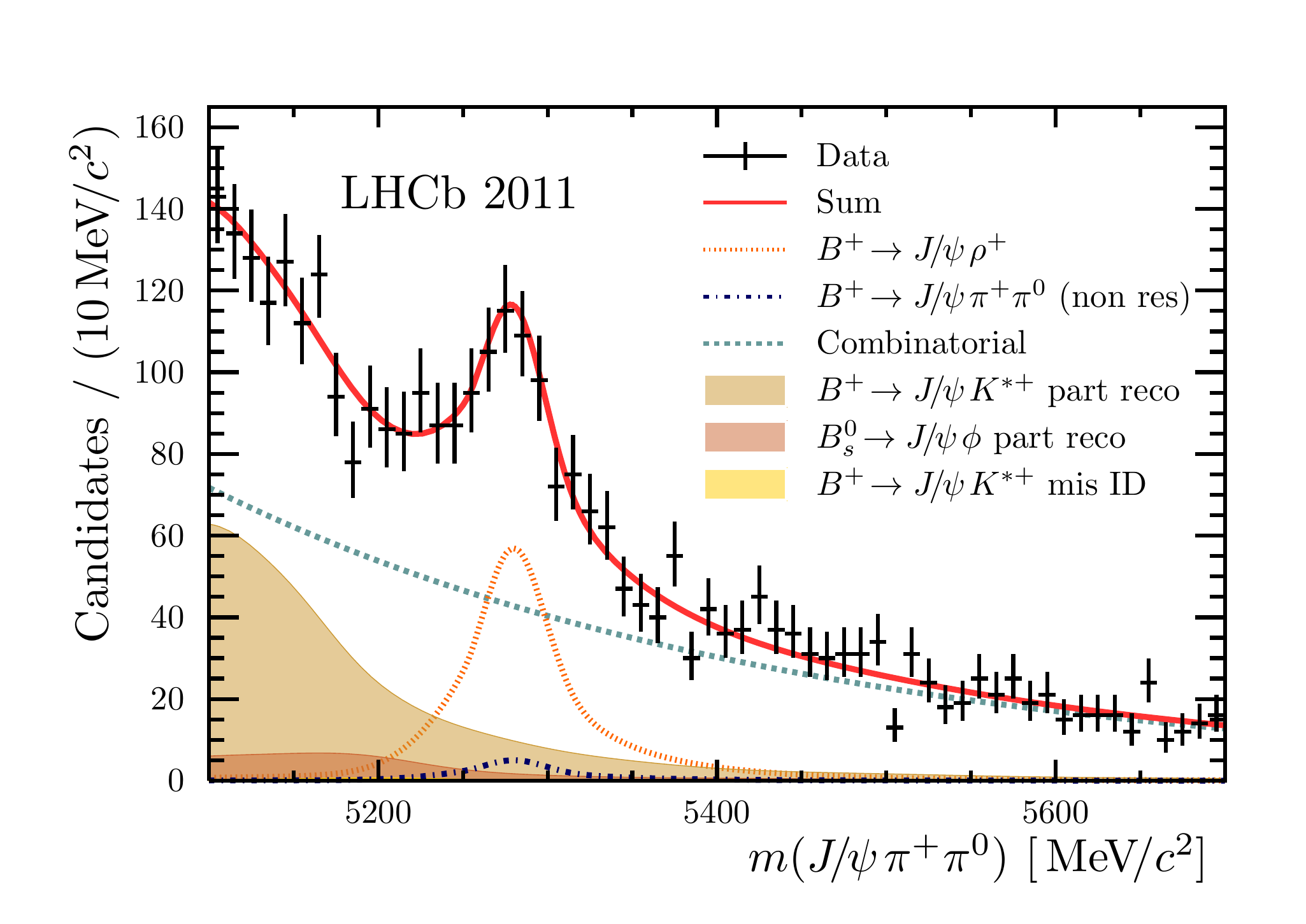}
    \includegraphics[width=0.49\linewidth]{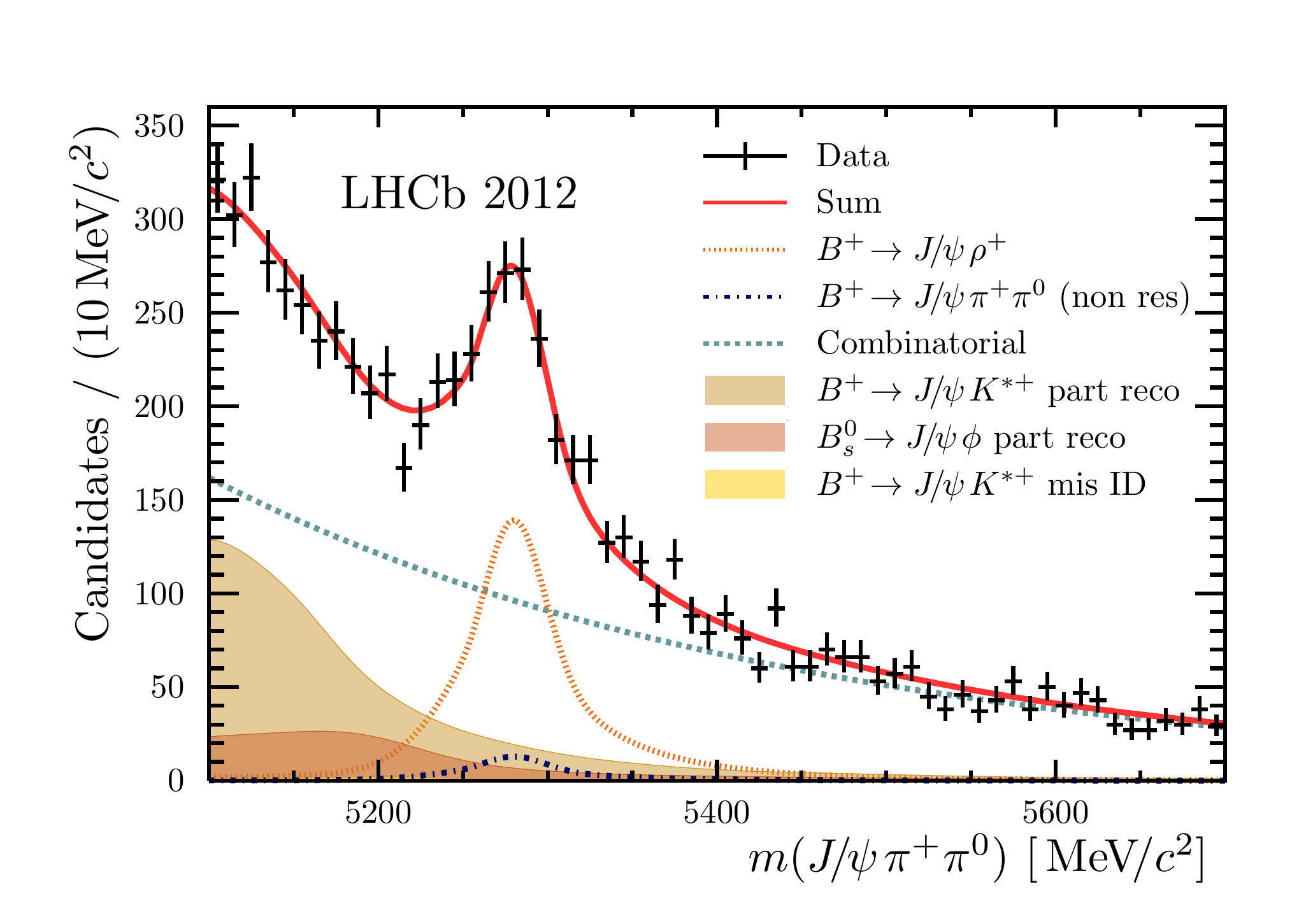}
    \vspace*{-0.5cm}
  \end{center}
  \caption{The \jpsi\pip\piz invariant mass distributions for the (left) 2011 and (right) 2012 data sets. In the legend, ``non res'' stands for non-resonant background, ``part reco'' for partially reconstructed background and ``mis ID'' for background involving a misidentification of a kaon. The mis ID background is small and not visible in these figures.}
  \label{fig:BMass}
\end{figure}

\begin{figure}[tb]
  \begin{center}
    \includegraphics[width=0.49\linewidth]{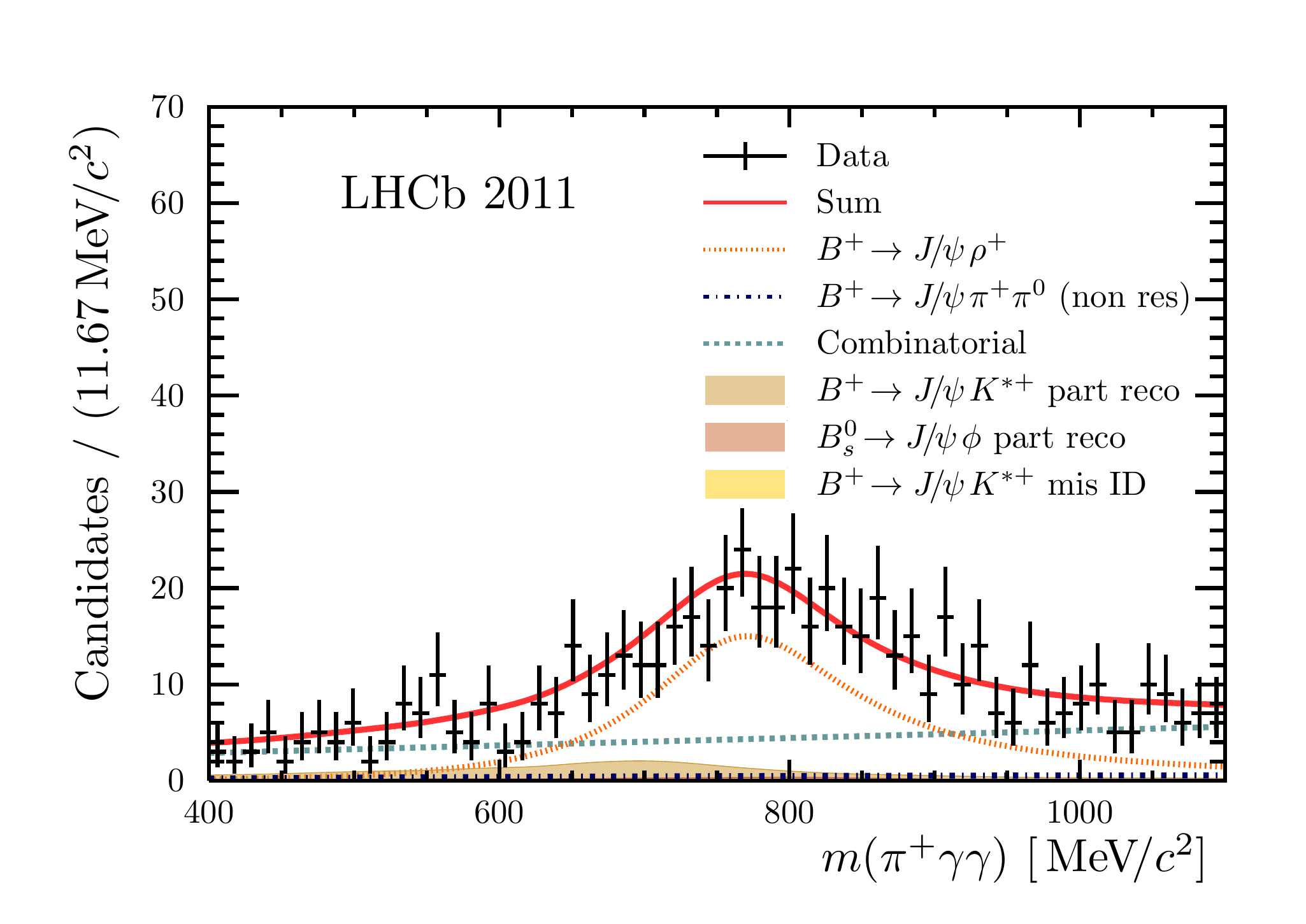}
    \includegraphics[width=0.49\linewidth]{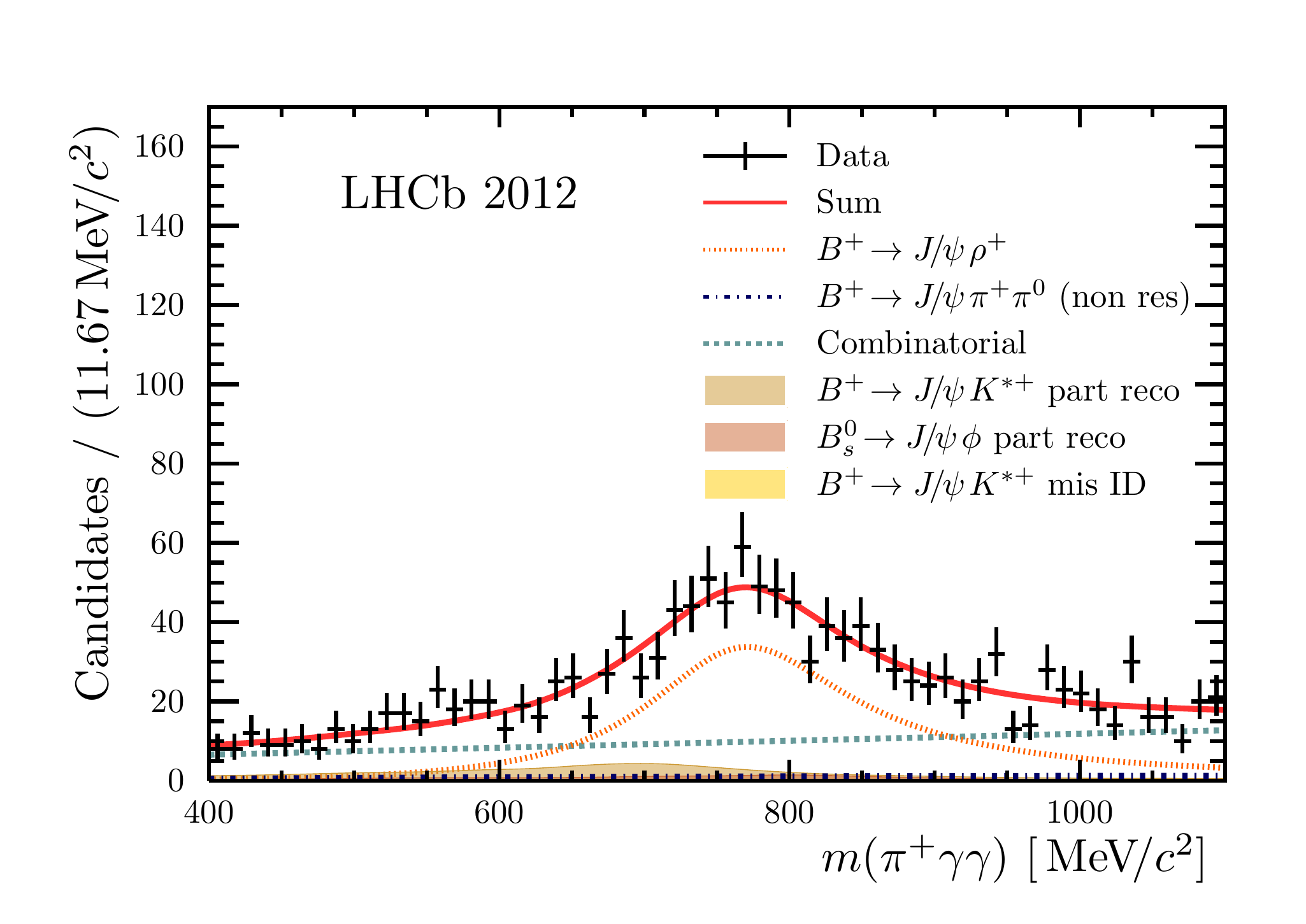}
    \vspace*{-0.5cm}
  \end{center}
  \caption{The \pip\g\g invariant mass distributions for the (left) 2011 and (right) 2012 data sets for $m_{\jpsi\pip\piz}$ between 5250\mevcc and 5310\mevcc. The part reco and mis ID backgrounds are small in the given mass region and not visible in these figures.}
 \label{fig:rhoMass}
\end{figure}

The one-dimensional projections of the two-dimensional simultaneous fit to the 2011 and 2012 data set are shown in Fig.~\ref{fig:BMass} for $m_{\jpsi\pip\piz}$  and in Fig.~\ref{fig:rhoMass} for $m_{\pip\g\g}$, where only events between 5250\mevcc and 5310\mevcc in $m_{\jpsi\pip\piz}$ are considered. 
For the full fit regions in $m_{\jpsi\pip\piz}$ and $m_{\pip\g\g}$ a total of $489\pm32$ ($1090\pm70$) signal decays are observed in 2011 (2012).
The fraction of \decay{\Bu}{\jpsi \pip\piz} decays that do not proceed via the $\rho^{+}$ resonance is measured to be 
$(8.4^{+6.1}_{-6.2})$\% 
in the $m_{\pip\g\g}$ interval  from 400\mevcc to 1100\mevcc. 

The value of \ACP is calculated using
\begin{equation}
\ACP = \mathcal{A}^{\CP}_\text{raw} - \mathcal{A}^\text{prod},
\end{equation}
where $\mathcal{A}^{\CP}_\text{raw}$ is the raw asymmetry, determined from a fit to the 2011 and 2012 data sets, split by the charge of the \B meson, where all fit components are modelled  as in the branching fraction measurement; and $\mathcal{A}^\text{prod}$ is the production asymmetry of \Bu mesons in \lhcb. For 2011 and 2012 they were measured to be ($-0.41 \pm 0.49 \pm 0.11$)\% and ($-0.53 \pm 0.31 \pm 0.10$)\%, respectively~\cite{LHCb-PAPER-2016-054}, with the first uncertainty being statistical and the second systematic.

For the background contribution from the \BuToJPsiKstp decay with \decay{\Kstarp}{\KS\pip}, the \CP asymmetry is fixed to the known value of ($-4.8\pm3.3$)\%\cite{PDG2018} after correcting for the production asymmetry. For the partially reconstructed background from \BsToJPsiPhi decays, no charge asymmetry is expected, as the final state is identical for both \Bs and \Bsb mesons.

\section{Systematic uncertainties}
\label{sec:systematics}

\subsection{Uncertainties on the branching fraction}

The systematic uncertainties  for the branching fraction measurement are summarized in Table~\ref{tab:brSystematics}.
The trigger efficiencies are derived from simulation. The ratio of efficiencies for \BuToJPsiKp and \BuToJPsiRhop decays has a small deviation from unity due to the slightly different kinematical distributions of the \jpsi mesons. To account for potential mismodelling of the impact of the \Kp and \rhop on the \jpsi trigger efficiency in simulation, the \pt distributions of the \Bu and \jpsi mesons in \BuToJPsiKp decays are weighted to match those in \BuToJPsiRhop decays, and the trigger efficiency is reevaluated. The resulting difference between this ratio of trigger efficiencies and unity is taken as a systematic uncertainty.
 
For the charged-particle reconstruction efficiency, the correction factors between simulation and data are varied within their uncertainties, and the effect on the ratio of \BuToJPsiRhop and \BuToJPsiKp decays is evaluated. The uncertainties consist of both statistical and systematic components, where the latter are due to the limited precision on the knowledge of the \lhcb material budget. Only the contribution of the material budget uncertainty and the different interaction cross-section of pions and kaons with the material significantly contribute to the uncertainty.

The uncertainty on the \piz reconstruction efficiency is dominated by the uncertainty on the branching fraction of the \BuToJPsiKstp decay, which is used in its calculation, and the number of \BuToJPsiKstp candidates in each kinematic bin. These values are added in quadrature to the systematic uncertainty of the method.

The systematic uncertainty for the charged-pion identification efficiency is evaluated by calculating the difference in efficiency between the nominal method, which is performed in intervals of \pt, pseudorapidity, and the number of tracks in the event, and an alternative, unbinned method. In addition, a different quantity for the event multiplicity and a different scheme of interval boundaries are used. All three uncertainties are added in quadrature. A similar procedure is used for the kaon and muon identification efficiencies, resulting in smaller systematic uncertainties.

For \BuToJPsiKp decays, the uncertainty on the selection efficiency is dominated by the limited size of the simulated data set. For the \BuToJPsiRhop selection, the largest component of this uncertainty arises from potential discrepancies in the \piz identification variable between simulation and data. The decay \BuToJPsiKstp is used to weight the distribution of this variable, and the change in the multivariate classifier efficiency with respect to the baseline value is taken as a systematic uncertainty. A smaller contribution comes from the limited size of the simulated data set. 

The uncertainty on the procedure to remove multiple-candidate events in the ${\BuToJPsiRhop}$ channel is determined on simulation by calculating the efficiency to select events that contain only one candidate, where the simulated events are weighted so that the event occupancy matches that observed in data. In addition, the invariant mass distributions in data are fitted for all events and events that only contain one candidate, where the signal shape is unchanged for both fits, {\em i.e.}\ the peaking background arising when the charged particles from a genuine signal decay are combined with a prompt \piz meson is ignored. The difference in efficiency between data and simulation 
is taken as a systematic uncertainty.

The two-dimensional kernel density estimators use adaptive kernel widths. The effect of the kernel width is tested by setting it to a constant value, either higher or lower than the default value, and taking the resulting difference in the branching fraction as a systematic uncertainty.
The tail parameters of the Crystal Ball function in \BuToJPsiRhop decays are varied by $\pm$20\% with respect to the baseline values to account for the uncertainties in the fit to simulation.  The value of $m_{\rhop}$ of the Breit--Wigner function is left free to vary, instead of being fixed to its baseline value. 
Furthermore, to take a possible difference in the experimental resolution of the $\rho^{+}$ mass between data and simulation into account, the value of $\Gamma_{0}$ of the Breit--Wigner shape is altered by the difference in the width of the \Bp mass peak between data and simulation in the decay \BuToJPsiKstp.
All differences with respect to the branching fraction of the baseline fit are added in quadrature and added to the overall systematic uncertainty.
To take possible correlations in the signal shape between $m_{\jpsi\pip\piz}$ and $m_{\pip\g\g}$ into account, the baseline model is replaced by a two-dimensional kernel density estimator whose shape is obtained from simulation. The resulting difference in the branching fraction with respect to the baseline model is taken as a systematic uncertainty.

The polarization of the decay products of the \Bu meson in the \BuToJPsiRhop decay is unknown. Using simulated decays where $L_{\text{eff}}$ is set to 1 or 2, the branching fraction is recalculated and the difference in the observed branching fraction with respect to the baseline result is taken as a systematic uncertainty. As a consistency test, simulated decays with $L_{\text{eff}} = 0$ are studied. A small bias with respect to the baseline branching fraction is observed, which is corrected for and taken as a systematic uncertainty.

To test the effect of different polarization amplitudes of \BuToJPsiRhop in simulation and data on the efficiency to reconstruct the decay, the values of the helicity amplitudes from \decay{\Bd}{\jpsi\rho^0}\cite{LHCb-PAPER-2014-012} are used in the simulation of \BuToJPsiRhop decays and varied within their uncertainties. The largest deviation with respect to the baseline value is taken as a systematic uncertainty.

In order to estimate the systematic uncertainty from the chosen fit range, the fit to determine the branching fraction is repeated 1000 times with random intervals in $m_{\jpsi\pip\piz}$ and $m_{\pip\g\g}$ larger or smaller than those of the baseline fit. The width of the resulting distribution of the measured branching fractions of \BuToJPsiRhop decays for each interval is taken as a systematic uncertainty. A small bias with respect to the baseline result was observed. The bias is corrected for and also taken as a systematic uncertainty.

As an alternative modelling for the nonresonant contribution in $m_{\pip\g\g}$, the shape is modelled with the form $(P_{\jpsi}/m_{B})^{2}$\cite{LHCb-PAPER-2012-045}, with $m_{B}$ the known mass of the \Bu meson\cite{PDG2018}. The difference of the branching fraction with respect to the baseline fit is taken as a systematic uncertainty.

A contribution from interference between the nonresonant \decay{\Bu}{\jpsi\pip\piz} and signal \BuToJPsiRhop decays could arise due to an asymmetric efficiency of the angular distribution of the \pip\piz system. To assess the effect of not taking the interference term in the branching fraction fit into account, several samples of one million simulated decays are generated including the interference term, where for each sample a different fixed phase for the nonresonant contribution is chosen. The shape of the angular acceptance is taken from the full simulation of \BuToJPsiRhop decays. Each sample is fitted with the baseline description, which does not include the interference term. The largest relative difference in the signal yield with respect to the generated value is taken as a systematic uncertainty. To investigate possible exotic resonance contributions, the invariant masses of \jpsi\pip and \jpsi\piz are inspected and no excess compared to the expectation is found.

Given the nature of the systematic uncertainties on the \piz and charged-particle reconstruction efficiencies, the selection efficiency and the removal of multiple candidates, their correlation in 2011 and 2012 is set to 1. 
All other uncertainties either result from a common fit to the combined data sets of 2011 and 2012 or are treated as uncorrelated.

\begin{table}[t]
\caption{Systematic uncertainties on the branching fraction \BR(\BuToJPsiRhop).} 
\begin{center}
\begin{tabular}{l | c }
Source of uncertainty & Relative uncertainty [\%] \\
 \hline    
Trigger efficiency & 1.4  \\ 
Charged particle reconstruction efficiency & 0.5  \\
\piz reconstruction efficiency &  6.3  \\ 
Hadron identification efficiency    &  2.1  \\ 
Muon identification efficiency   &  0.4  \\ 
Selection efficiency \BuToJPsiKp &  0.1 \\ 
Selection efficiency \BuToJPsiRhop & 1.8 \\
Removal of multiple candidates & 1.2 \\
Fit function  & 4.0 \\
\BuToJPsiRhop polarization & 2.2 \\
Fit ranges               & 1.6\\
Nonresonant line shape &  1.5\\	
Neglecting interference & 2.8\\
 \hline
Quadratic sum         & 9.1\\
 \end{tabular}
 \label{tab:brSystematics}
 \end{center}
 \end{table}

\subsection{Uncertainties on the \boldmath{\CP} asymmetry}
Most systematic uncertainties cancel when calculating the \ACP ratio. The remaining contributions are listed in Table~\ref{tab:acpSystematics}. The largest contributions come from the uncertainty on the knowledge of the direct \CP asymmetry of the \BuToJPsiKstp decay for the partially reconstructed background, and the limited knowledge of the production asymmetry for \Bu mesons in the 2011 and 2012 data sets.
The signal model is again replaced by a two-dimensional kernel density estimator to take possible correlations into account, taking the difference in the \CP-asymmetry results as a systematic uncertainty.
The ratio of the positive and negative pion detection efficiencies $\varepsilon (\pip) / \varepsilon (\pim)$ has been measured at the \lhcb experiment~\cite{LHCb-PAPER-2012-009} and found to be compatible with unity over a broad range of momenta and transverse momenta, with an uncertainty of about 0.5\%. This uncertainty is added as a systematic uncertainty for \ACP.

\begin{table}[t]
\caption{Systematic uncertainties on the direct \CP asymmetry of \BuToJPsiRhop decays.}
\begin{center}
\begin{tabular}{l | c }
Source of uncertainty & Uncertainty  \\
 \hline    
\Bu production asymmetry and background asymmetry & 0.006 \\ 
Signal fit function & 0.005  \\
Pion detection asymmetry & 0.003  \\
 \hline
Quadratic sum         & 0.008 \\
 \end{tabular}
 \label{tab:acpSystematics}
 \end{center}
 \end{table}

To evaluate a possible bias on the asymmetry result, a permutation test is performed where the data set is split 1000 times randomly, instead of by the charge of the \Bp meson, and the asymmetry is evaluated. As an additional check,  simulated decays are added to each of these randomly split samples, corresponding to $\pm 5$\% and  $\pm 10$\% asymmetry, to assess the robustness of the asymmetry fit for a non-zero \ACP value. No bias in the resulting distributions is observed.
The total systematic uncertainty for \ACP is formed by adding all individual components in quadrature.

\section{Results and Summary}

Using Eq.~\eqref{eq:master} the ratio of the branching fractions of the decays \BuToJPsiRhop and \BuToJPsiKp is determined to be  
\[
\frac{\BR(\BuToJPsiRhop)}{\BR(\BuToJPsiKp)} = 0.0378 ^{+0.0025}_{-0.0024}
\pm 0.0035
\]
in a combined fit to the 2011 and 2012 data sets, where the first uncertainty is statistical and the second systematic. The ratio of efficiencies $\frac{\varepsilon_{\BuToJPsiKp}}{\varepsilon_{\BuToJPsiRhop}}$ is $27.2\pm 2.0$ for the combination of both data sets, it is dominated by the low efficiency to select a \piz with a sufficiently high transverse momentum.
The branching fraction for the decay \BuToJPsiRhop is

\[
\BR(\BuToJPsiRhop) = (3.81 ^{+0.25}_{-0.24} \pm 0.35) \times 10^{-5}. 
\]
The \CP asymmetry is measured to be
\[
\ACP(\BuToJPsiRhop) = -0.045^{+0.056}_{-0.057} \pm 0.008.
\]
Both results are the most precise to date and are consistent with previous measurements. 
Furthermore, the measured value of \ACP is consistent with the corresponding measurement using \BdToJPsiRhoz decays, as expected from isospin symmetry~\cite{LHCb-PAPER-2014-058}.

\section*{Acknowledgements}
%
%
\noindent We express our gratitude to our colleagues in the CERN
accelerator departments for the excellent performance of the LHC. We
thank the technical and administrative staff at the LHCb
institutes.
We acknowledge support from CERN and from the national agencies:
CAPES, CNPq, FAPERJ and FINEP (Brazil); 
MOST and NSFC (China); 
CNRS/IN2P3 (France); 
BMBF, DFG and MPG (Germany); 
INFN (Italy); 
NWO (Netherlands); 
MNiSW and NCN (Poland); 
MEN/IFA (Romania); 
MSHE (Russia); 
MinECo (Spain); 
SNSF and SER (Switzerland); 
NASU (Ukraine); 
STFC (United Kingdom); 
NSF (USA).
We acknowledge the computing resources that are provided by CERN, IN2P3
(France), KIT and DESY (Germany), INFN (Italy), SURF (Netherlands),
PIC (Spain), GridPP (United Kingdom), RRCKI and Yandex
LLC (Russia), CSCS (Switzerland), IFIN-HH (Romania), CBPF (Brazil),
PL-GRID (Poland) and OSC (USA).
We are indebted to the communities behind the multiple open-source
software packages on which we depend.
Individual groups or members have received support from
AvH Foundation (Germany);
EPLANET, Marie Sk\l{}odowska-Curie Actions and ERC (European Union);
ANR, Labex P2IO and OCEVU, and R\'{e}gion Auvergne-Rh\^{o}ne-Alpes (France);
Key Research Program of Frontier Sciences of CAS, CAS PIFI, and the Thousand Talents Program (China);
RFBR, RSF and Yandex LLC (Russia);
GVA, XuntaGal and GENCAT (Spain);
the Royal Society
and the Leverhulme Trust (United Kingdom);
Laboratory Directed Research and Development program of LANL (USA).





\addcontentsline{toc}{section}{References}
\setboolean{inbibliography}{true}
\bibliographystyle{LHCb}
\bibliography{main,LHCb-PAPER,LHCb-CONF,LHCb-DP,LHCb-TDR}
 
 \newpage

\clearpage 
\newpage

\centerline{\large\bf LHCb collaboration}
\begin{flushleft}
\small
R.~Aaij$^{28}$,
C.~Abell{\'a}n~Beteta$^{46}$,
B.~Adeva$^{43}$,
M.~Adinolfi$^{50}$,
C.A.~Aidala$^{78}$,
Z.~Ajaltouni$^{6}$,
S.~Akar$^{61}$,
P.~Albicocco$^{19}$,
J.~Albrecht$^{11}$,
F.~Alessio$^{44}$
% ...

{\footnotesize \it
% ...
$ ^{6}$Clermont Universit{\'e}, Universit{\'e} Blaise Pascal, CNRS/IN2P3, LPC, Clermont-Ferrand, France\\
% ...
$ ^{11}$Fakult{\"a}t Physik, Technische Universit{\"a}t Dortmund, Dortmund, Germany\\
% ...
$ ^{19}$INFN Laboratori Nazionali di Frascati, Frascati, Italy\\
% ...
$ ^{28}$Nikhef National Institute for Subatomic Physics, Amsterdam, Netherlands\\
% ...
$ ^{43}$Instituto Galego de F{\'\i}sica de Altas Enerx{\'\i}as (IGFAE), Universidade de Santiago de Compostela, Santiago de Compostela, Spain\\
$ ^{44}$European Organization for Nuclear Research (CERN), Geneva, Switzerland\\
% ...
$ ^{46}$Physik-Institut, Universit{\"a}t Z{\"u}rich, Z{\"u}rich, Switzerland\\
% ...
$ ^{50}$H.H. Wills Physics Laboratory, University of Bristol, Bristol, United Kingdom\\
% ...
$ ^{61}$University of Cincinnati, Cincinnati, OH, United States\\
% ...
$ ^{63}$Syracuse University, Syracuse, NY, United States\\
% ...
$ ^{78}$University of Michigan, Ann Arbor, United States, associated to $^{63}$\\
% ...
}
\end{flushleft}

\end{document}